# Development of an underwater inducive coupling communication system with power carrier technology

Author：Zhongxing Zhang

Abstract：Inductive coupling communication is one of the main methods of underwater communication systems due to its excellent comprehensive performance. However, the data transmission distance and operational power consumption need to be further enhanced. In this paper, an underwater induction coupling communication scheme based on power carrier technology is proposed to improve the transmission speed and reduce the bit error rate. The microcontroller of STM32L series with ultra-low power consumption was employed as the core of the system. Through the construction and simulation of the communication channel, the optimal parameters were determined. According to the circuit model of the power carrier communication, the effect of different modulation and demodulation methods to the signal transmission quality were discussed, which demonstrates the superiority of Differential Phase Shift Keying (DPSK). With the system-level low power design and onboard communication quality optimization, the device was developed. The test results in the laboratory environment show that the system can achieve efficient data communication with a rate of 115200bps and static power consumption as low as 660μA in the 700m channel. This study provides a practical design approach for the high-speed communication and Low-power operation of underwater communication systems.

Keywords：Underwater communication; Power carrier; Inductive coupling; High speed; Low power design

# 1 Introduction

Ocean monitoring and marine resource development are directly related to the sustainable development of human beings in the future [1]. The establishment of marine observation systems to timely acquire various ocean data for analyzing and processing plays an important role in alleviating resource shortage and regulating global climate [2]. Therefore, a high-speed, stable and long-life underwater data transmission system is crucial to the operation of marine monitoring platforms. Inductive coupling communication is one of the mainstream methods of marine monitoring due to its advantages of stable transmission, fast transmission speed and less susceptibility to environments [3]. However, this communication method exists serious defects of signal attenuation in long-distance transmission.

Researchers have conducted a great deal of investigations on anchor-coupled inductive coupling data transmission systems. For example, the inductive coupling communication system designed by Tianjin University [4, 5] can mount 10 sets of sensors in experiments, with a measurement depth of 1000m and a data transmission rate of 4800bps. The inductive coupling chain developed by the National Ocean Technology Center (NOTC) can accommodate 36 sets of sensor devices with its measure depths up to 4000m, compatible with both seawater and freshwater environments [6, 7]. Companies such as US Seabird [8], Canada's Abilene Marine Instruments [9], Japan's Alec Electronics have carried out extensive research and developed anchor-coupled inductive coupling data transmission products. The relevant parameters of the equipment developed by various organizations are shown in Table 1. Although these products have achieved high levels regarding measurement depth and number of sensors mounted, the signal transmission speed needs to be enhanced.

Table 1 Parameters of various induction coupling equipment.

| Institution | Number of sensors mounted | Modulation mode | Measure depth(m) | Transmission speed(bps) |
|---|---|---|---|---|
| Sea-Bird Scientific | 100 | DPSK | 7000 | 9600 |
| RBR | 32 | DPSK | 4000 | 4800 |
| NOTC | 36 | DPSK | 4000 | 4800 |
| Tianjin University | 10 | DPSK | 1000 | 4800 |
| Harbin Engineering University | - | CPM | 200 | 9600 |

Power carrier technology, superimposing low-frequency data signals onto high-frequency carrier signals [10], is widely used in smart home [11], communication [12] and other fields. In this paper, power carrier technology is combined with inductive coupling communication. The optimal channel parameters are determined by constructing an inductive coupling communication channel model and analyzing the influence of channel materials and parameters on signal transmission quality. In order to reduce the static power consumption of the system, this paper proposes a system-level low-power scheme and a low-power standby mode, and then designs an underwater low-power communication system based on STM32L ultra-low-power microprocessor. According to the system communication quality requirements, the hardware and software design is carried out, and the data signal is carrier modulated using power carrier technology to improve the signal transmission speed. Finally, single-point and multi-point communication tests were carried out using the prototype in the laboratory environment to verify the low-power performance and high-speed communication performance of this system, which provides a theoretical reference and feasible solution for the design of underwater high-speed inductive coupling communication.

## 2 System design scheme

A typical underwater induction coupled communication system consists of an induction coupled channel, a surface control equipment, and underwater acquisition equipment [13], as shown in Fig. 1. Among them, the inductive coupling channel can be equivalent to a single-turn loop composed of seawater and cable. The data collected by the underwater equipment is coupled into the channel for transmission through electric-magnetic signal conversion, and the magnetic ring connected to the surface control unit realizes the signal conversion in the primary winding to complete the signal reception.

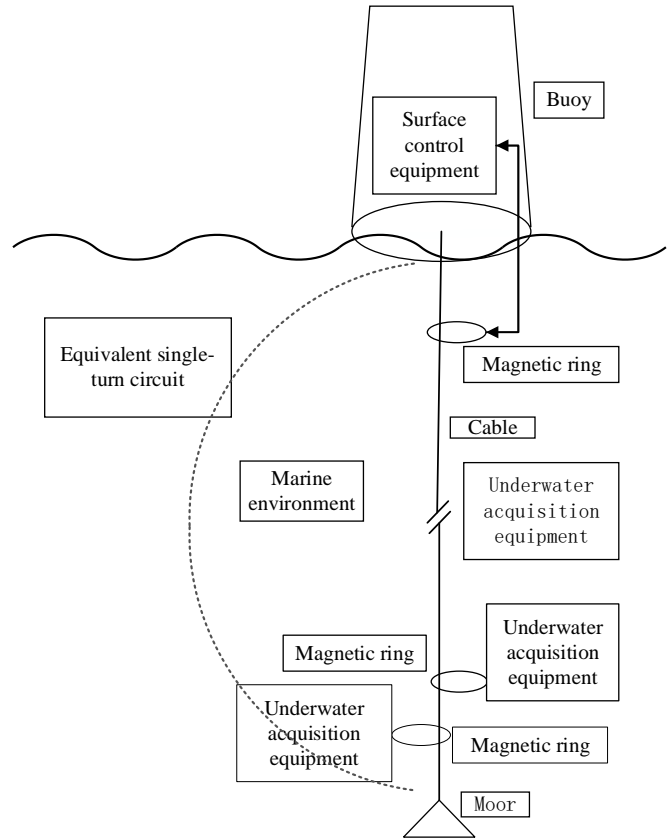

Fig. 1. Block diagram of a typical underwater inductive coupling data transmission system.

According to the communication principle of underwater inductive coupling, the power carrier technology was introduced to the system illustrated in Fig. 1, where the new configuration includes power conversion unit, sensors, control unit power carrier unit, and so on, as depicted in Fig. 2. The power conversion unit provides 3.3V and 12V voltage for the control unit, carrier communication unit, etc. The sensor unit sends the hydrological information collected by different sensors to the control unit for processing and transmission. The functions of the control unit include low-power state switching, signal transmission and reception, peripheral control, etc. The signal processing unit filters and amplifies the received signal to ensure the filtering of noise signals and the reception of instruction signals, thereby preventing false wake-up in low-power mode.

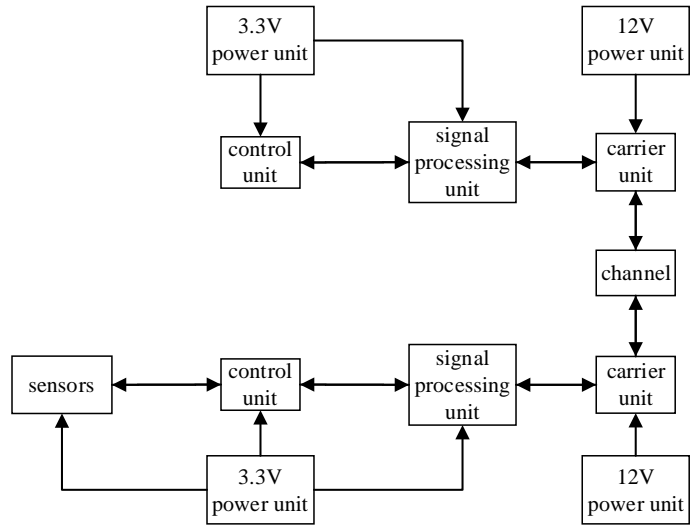

Fig. 2. Overall system structure.

## 2.1 Inductive coupling Channel Design and Analysis

The structure of a typical inductive coupling communication channel consists of three parts: coupling magnetic ring, cable, and coil [14]. The permeability, cross-sectional area, magnetic circuit length, and other factors of the magnetic ring material affect the magnitude of the inductance in the coil. In the induction coupled communication channel, the single turn loop formed by cables and seawater has a strong absorption and attenuation effect on the signal. In order to ensure the quality of the signal during transmission, it is necessary to choose magnetic ring materials with high magnetic permeability, low coercivity, high resistivity, and high operating frequency. According to the above parameters, compare various magnetic ring materials and choose manganese-zinc ferrite as the magnetic ring material, as shown in Fig. 3.

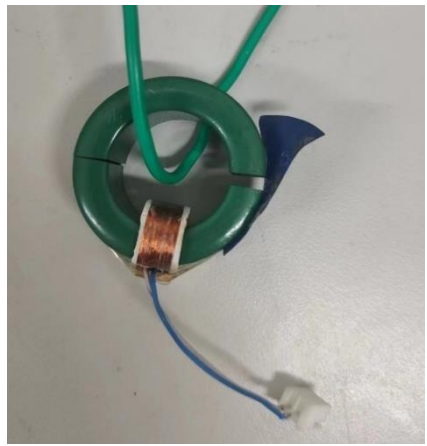

Fig. 3. Half circular magnetic ring.

When the frequency and amplitude of the signal are constant, increasing the number of coil turns will reduce the induced electromotive force in a single turn circuit, which is not conducive to signal transmission. Reducing the number of coil turns will lead to a decrease in the inductance value of the primary winding and a decrease in the induced electromotive force. Therefore, the appropriate number of turns of the coil can effectively improve the quality of signal transmission, this topic through a number of experiments to compare the quality of data transmission under different number of turns, the results are shown in Table 2. By comparing the data transmission quality under different turns, a coil turn of 4 is selected.

Table 2 Transmit signal and receive signal amplitude for different number of turns of coil.

| Turns ratio | Transmitter signal amplitude(V) | Receiver signal amplitude(mV) |
|---|---|---|
| 2：1：2 | 12.0 | 264 |
| 3：1：3 | 12.0 | 284 |
| 4：1：4 | 12.0 | 392 |
| 5：1：5 | 12.0 | 308 |
| 6：1：6 | 12.0 | 296 |
| 7：1：7 | 12.0 | 296 |
| 8：1：8 | 12.0 | 260 |

Underwater cables are mainly used to form a single turn loop with seawater. Considering that the application cable is often in a seawater environment and needs to have good corrosion resistance, wear resistance, and tensile strength, the 700m wrapped plastic steel cable in the research group is selected as the channel cable for underwater induction coupling communication, as depicted in Fig. 4.

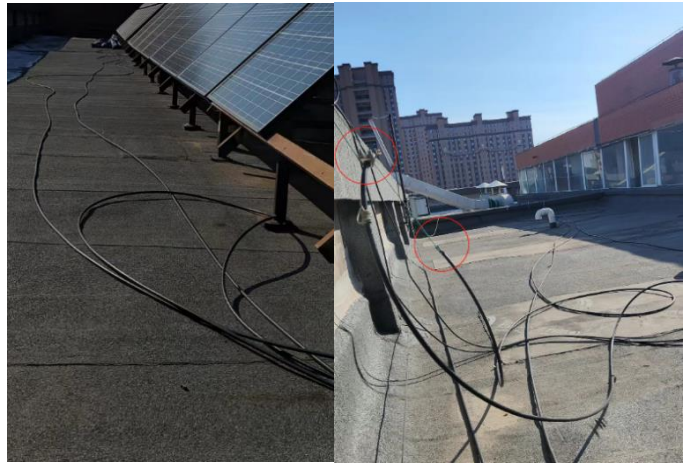

Fig. 4. Communication cable.

## 2.2 Hardware Circuit Design

In order to achieve long-term effective operation of the induction coupled underwater data transmission system under the influence of harsh environments such as ocean currents and ocean currents, and to reduce the probability of system damage, the hardware circuit of this system was designed with the goals of low power consumption, stable structure, and fast transmission rate, as depicted in 5.

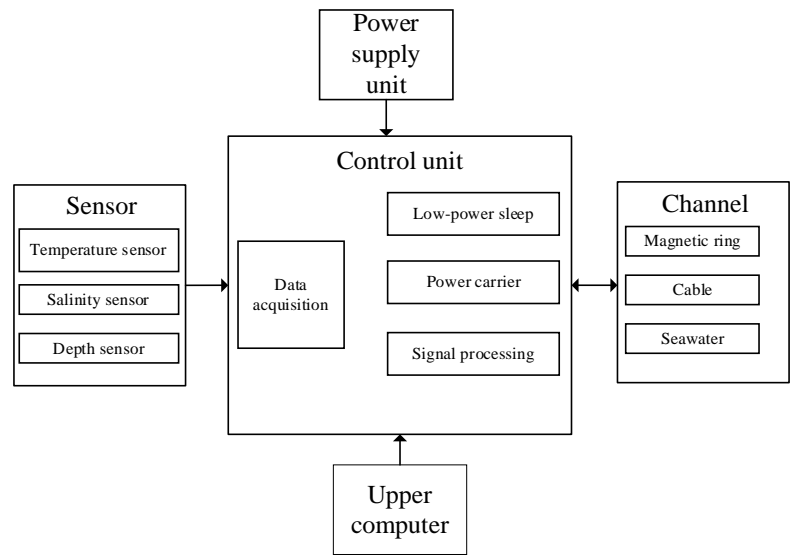

Fig. 5. Structure of underwater induction coupled communication system

Power management is the foundation of the entire system. All peripherals require power supply support. This device uses a 3.7V lithium battery as the power supply battery. The LDO module is designed to convert the input voltage into 3.3V voltage and provide it to the control unit, carrier communication, signal processing and other units. At the same time, the DC-DC module boosts the voltage from 3.7V to 12V and supplies it to the carrier signal amplification module of the power carrier unit, achieving the superposition and transmission of carrier signals and data signals.

The control unit is the core of the hardware system. According to the requirement of low-power consumption, this system chooses STM32L476 as the control chip. The chip is an ultra-low-power microcontroller based on the high-performance Arm Cortex-M4 core, operating at up to 80MHz with 48 pins, as shown in Fig. 6. The internal regulator can provide 1.71 to 3.6 V power supply and provide multiple working modes to adjust the system's working power consumption.

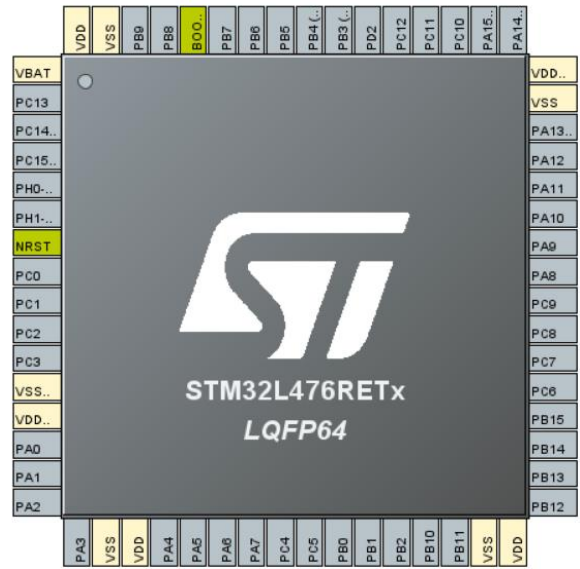

Fig. 6. Pinout diagram.

After receiving water command signals, the control unit controls the DHT11 sensor to collect temperature information. During testing, the control unit writes the temperature information collected into instructions and modulates it through the carrier communication unit. The modulated signal is sent to the channel for transmission.

Signal transmission in the channel there is attenuation and mixed noise problems, set the signal processing unit on the data filtering and amplification, to ensure that the control signal reception and noise filtering. Through the test, the system carrier frequency is 1.67MHz, the amplification is about 3 times, the gain bandwidth is about 5MHz, according to the gain bandwidth demand, choose LTC6261 chip to build the signal processing unit. Using an oscilloscope to observe the test results shown in Fig. 7. The unit can filter out low-frequency noise and amplify the received signal to the ideal threshold.

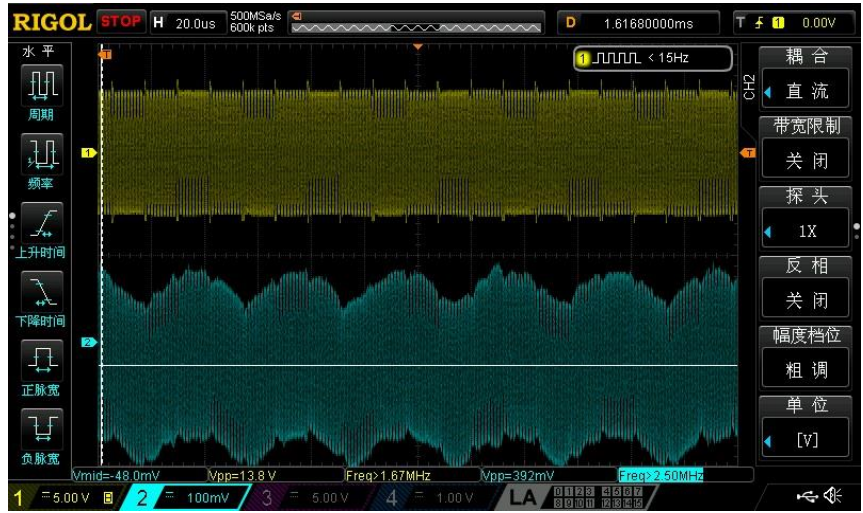

Fig. 7. LTC6261 Amplifier Test.

The host computer through the serial port control surface control equipment to send control commands and receive the hydrological data information. In the test phase, the host computer display can help system debugging, but also to make the signal reception has an intuitive form of expression as depicted in Fig. 8. Which 02, 49, 46, 68, 00, 53 for the sender's address, accept the data for 12, 34. D2, 68 for and check bits and heterodyne check bits, respectively.

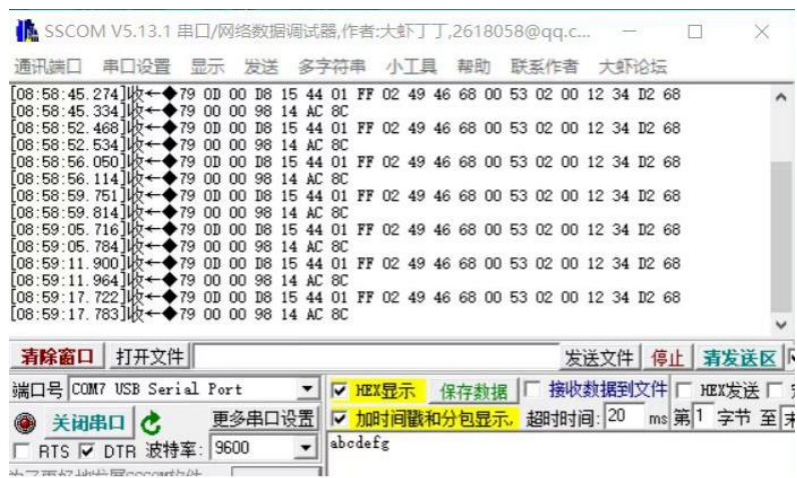

Fig. 8. Upper computer communication.

Due to underwater interference and signal attenuation, the data information collected by the sensor can not be transmitted in the baseband, it is necessary to carry out modulation processing, the system selects the DPSK modulation mode to modulate the signal as shown in Fig. 9, the modulation mode has a better anti-jamming ability, high-frequency transmission capability and does not produce reverse work effect.

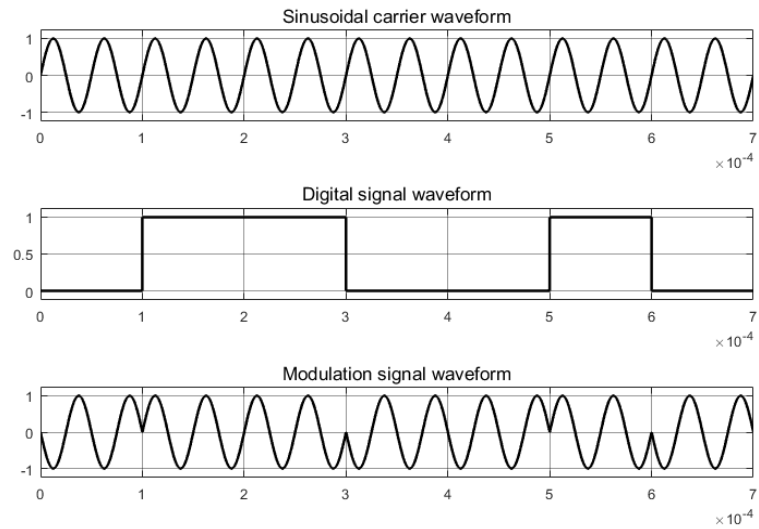

Fig. 9. Modulated waveform of DPSK signal.

The carrier unit superimposes a low-frequency data signal onto a high-frequency carrier signal for transmission. At the receiving side, the carrier unit is used to decode the modulated signal, and the data signal is obtained by removing the high-frequency carrier signal as depicted in Fig. 10. This system selects the SSC1642 carrier chip with high transmission rate to build the carrier communication unit.

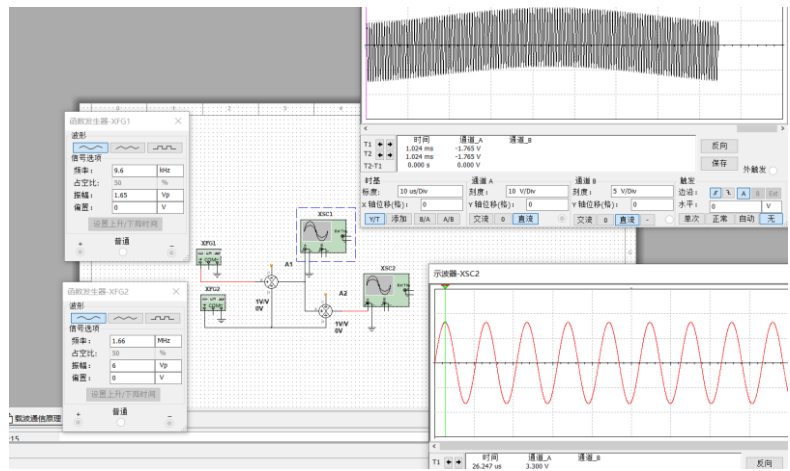

Fig. 10. Carrier signal realization principle and output signal.

According to the above design, using Altium Designer software on the schematic drawing, device selection and peripheral circuit design for each circuit unit, PCB design, printing and soldering are carried out according to the designed schematic. The prototype PCB board area of this device is 57mm*37mm, as shown in Fig. 11.

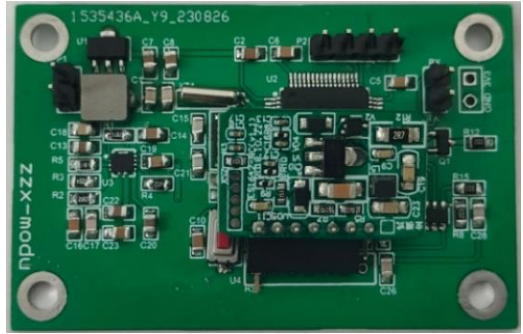

Fig. 11. Equipment prototype.

## 3 Software development

### 3.1 Software design

According to the STM32L476 controller chip, the program is developed using KEIL and STM32CubeMX, and the software flowchart is shown in Fig. 12.

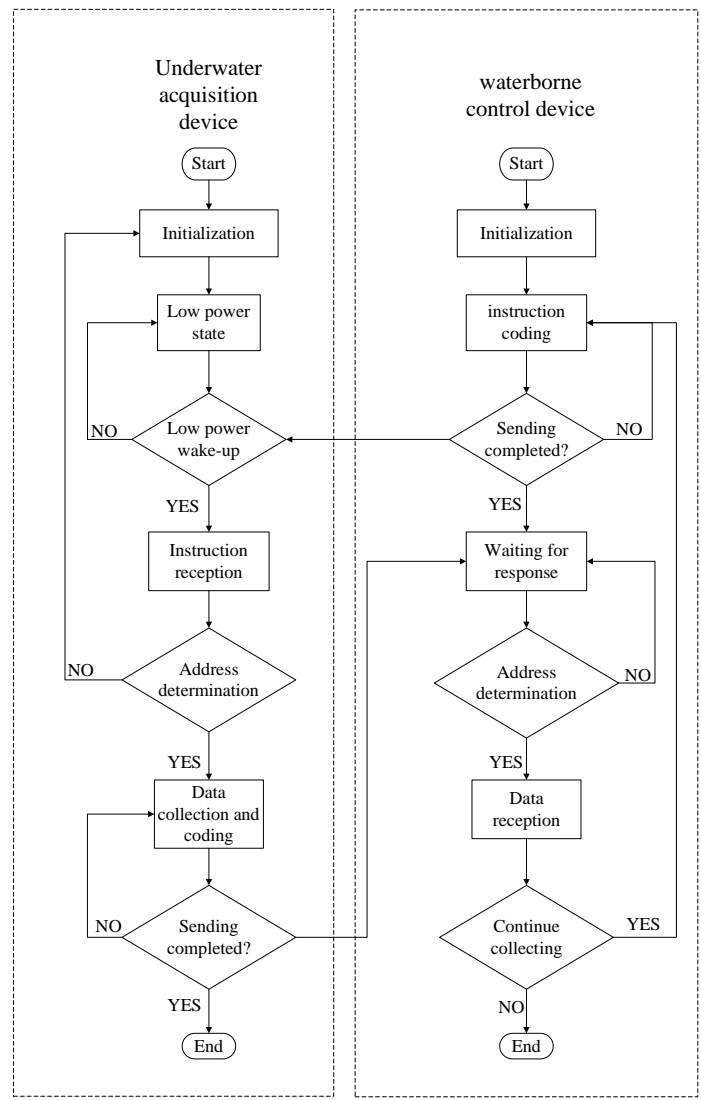

Fig. 12. Program flow.

The underwater acquisition device enters the standby state after system initialization, while the surface control device selects the underwater acquisition device to be woken up through the address. Underwater acquisition equipment from the magnetic ring to receive the command signal transmitted by the surface control equipment, the first address judgment, whether to respond, if the address does not match the device continues to standby. Otherwise, it will collect data and send it to the power carrier unit through the serial port. surface control equipment for underwater transmission of data signals address matching, decide whether to receive.

### 3.2 Working mode

Multiple operating modes of the system were tested for power consumption, as depicted in Table 3. After comparison, STOP1 mode was selected as the low-power state of the system.

Table 3 Power consumption comparison.

| Mode | Regulator | CPU | Clocks | Wakeup source | Consumption | Wakeup time |
| --- | --- | --- | --- | --- | --- | --- |
| RUN | MR | Yes | On | N/A | 12mA | N/A |
| LPRUN | LPR | Yes | Except PLL | N/A | 3.35mA | 64μs |
| Sleep | MR | No | All | Interrupt or event | 1.2mA | 6 cycles |
| STOP1 | LPR | No | LSE、LSI | UART, etc. | 566μA | 7.8μs |
| Shutdown | OFF | No | LSE | RTC | 0.23μA | 306μs |

In the RUN mode, the CPU running frequency can be designed according to the needs, each function module is in the configuration state, the CPU and peripherals are in the working state, and the device power consumption is the highest, about 10-12mA.

The LPRUN mode reduces the system power consumption by lowering the system main frequency to 2MHz. However, the mode must be entered from RUN mode, which does not meet the system-level low power operation mode requirement.

STOP1 mode uses a low-power regulator, and by shutting down peripherals such as the CPU and FLASH, the clock can only use a low-speed clock. As a result, the STOP mode achieves the lowest power consumption and preserves the contents of the

SRAM and registers. Different STOP modes have different wake-up sources, power consumption and wake-up time. Based on the above characteristics, this system selects STOP1 as the low-power state

### 3.3 Low-power management

Considering the poor economy of replacing batteries in the deep sea, this system proposes a power management scheme to reduce the static power consumption of the system by designing the system's work mode, power gating, and system main frequency.

(1) A variety of low-power operation modes provided by the control chip are tested, and STOP1 with low static power consumption, applicable wake-up source and short wake-up time, is selected as the low-power operation mode of this system. After entering the STOP1 mode, the system shuts down the CPU, clocks and peripherals and reduces the system operating frequency to reduce the static power consumption of the system, and sets the comparator interrupt to departure low-power wakeup.

(2) The GPIO provided by the control chip is used as the control signal output pin to disable the functional module, so that the functional module can be dormant in time to reduce the overall power consumption of the system and ensure the service life of the device. For example, the 12V power supply module enters the shutdown state when there is no need for data transmission.

(3) Design switching circuits for high power consumption modules to effectively reduce the static operating current of the circuit. Such as the static power consumption of the carrier unit is 0.045W, the carrier unit 3.3V pin to join the switching circuit, effectively reducing the system static power consumption.

## 4 Test and Analysis

Combined with channel design and hardware and software development, we have built a test environment for inductive coupling communication system, completed the development of prototypes for control equipment and acquisition equipment, and conducted system power consumption and multi-point communication testing.

## 4.1 Power consumption test

The low-power performance of the underwater acquisition device was tested in a laboratory test environment. By connecting the red pen of the power supply to the red pen of the multimeter, the negative pole of the power supply to the negative pole of the device, and the multimeter is turned to the current gear, and its black pen is connected to the positive input of the device, the total static operation power consumption of the system is tested as depicted in Fig. 13.

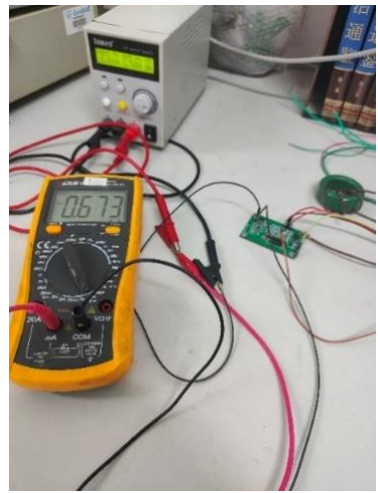

Fig. 13. System static power consumption test.

Power consumption tests are performed on the carrier communication, signal processing, and control units respectively. In the standby state, the power supply of the carrier communication unit is turned off, and the power consumption test of the remaining units is shown in Fig. 14. The static power consumption of the carrier communication unit is about 130μA.

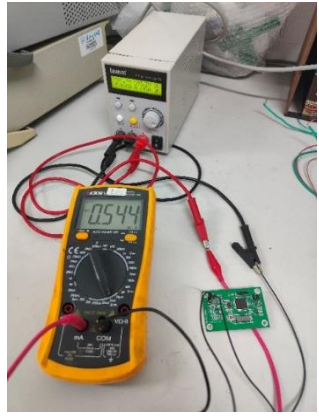
Fig. 14. Power consumption test without carrier unit.

In the same way, the signal processing and power conversion units are tested for power consumption, and it can be concluded that the static power consumption of the signal processing unit is about 300μA, the static power consumption of the power conversion unit is about 180μA, and the static power consumption of the master unit is about 50μA. The test results are close to the static power consumption given by the manuals of each chip.

## 4.2 Single-point communication test

In the single-point communication test, this paper first in the 2m cable constitutes the channel for test experiments as depicted in Fig. 15, the cable ends immersed in salt water to simulate the underwater equivalent, the control equipment and acquisition equipment through the magnetic ring socketed on the cable, in the 3.7V power supply for single-point communication test experiments. The control equipment sends the command signal with 12H and 34H as data acquisition commands, and the receiving equipment receives the signal and sends the signal with 00H and FFH as data signals.

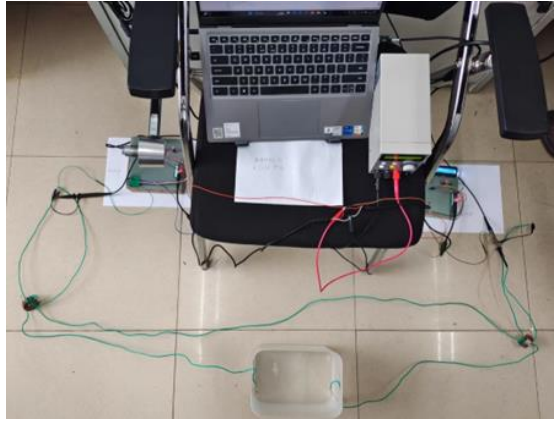

Fig. 15. Single-point communication test environment.

In the single-point communication test, the surface control equipment through the connection of the serial port will be data displayed as shown in Fig. 16. This system communication relay depth of 1 that underwater acquisition equipment and surface control equipment for direct transmission, underwater acquisition equipment address 64, 49, 46, 68, 00, 53, transmission data for 00, FF and test setup parameters are the same, the test proves that the feasibility of power carrier technology applied to underwater inductive coupling communication systems.

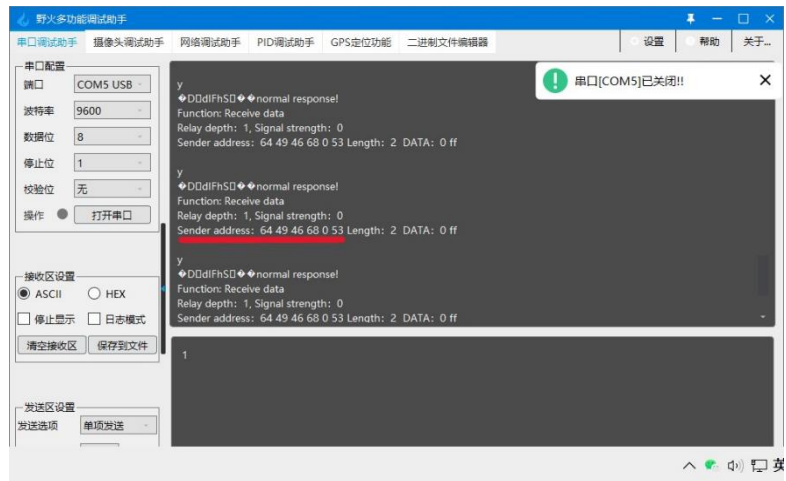

Fig. 16. Single-point communication test.

### 4.3 Multi-point communication test

Single-point test experiment verifies the communication function of this system, in order to realize the information acquisition at different depths, multiple underwater acquisition devices are installed at different depths of the 700m cable, this paper carries out the multi-point communication experiment of a single surface control device to five underwater acquisition devices in the laboratory environment by setting up different 6-byte addresses of different underwater acquisition devices, and the experimental environment is shown in Fig. 17.

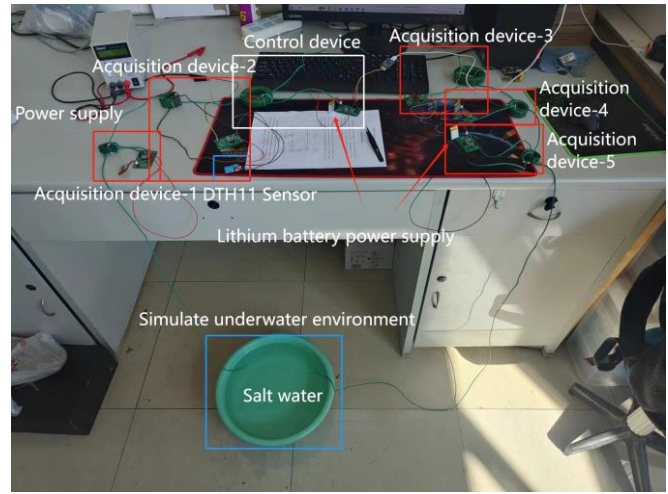

Fig. 17. Experiment system of multi-point communication.

In the experimental environment shown in Fig. 17, the surface control device selects the address of different underwater acquisition devices for command sending and data receiving, one of the underwater acquisition devices and the surface control device are connected to the PC through the serial line in order to observe the data sending and receiving situation, and the experimental results are shown in Fig. 18. The relay depth is 1, the signal strength is the highest, and the address of the underwater acquisition device is (89, 47, 46, 68, 00, 53), the length of the received data is 2, and the data is 00 ff.

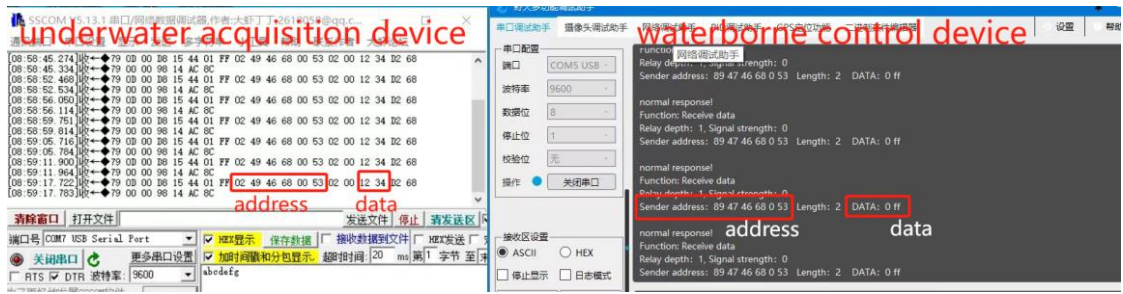

Fig. 18. Description of multi-point communication test.

The result matches with the setup instructions in the program, and the data are correct in several experiments, while the check bits set in the frame format prove the reliability of the data transmission. The experimental results show that the system can perform high quality signal transmission at 9600bps.

After completing the information monitoring test of the acquisition device and the control device, the serial port line of the underwater acquisition device is removed, and the programmed power supply is utilized for power supply and power consumption monitoring. In that experiment, the underwater acquisition device wakes up the surface control device with different addresses respectively, and the communication test results are shown in Fig. 19, the serial port baud rate is 115200bps, and it can be seen in the data received by the serial port assistant that the addresses are (03, 03, 46, 68, 00, 53), (11, 01, 46, 68, 00, 53), (39, 41, 46, 68, 00, 53), etc. The data received back from the underwater acquisition device is the temperature data collected by the DHT11, of which 13, 07 indicates that the temperature at that time was 19.7℃. The temperature data collected by the No.2 underwater acquisition device is 24.8℃ by enhancing the surrounding environment temperature through the hot blowing air. The multi-point test experiments can prove the high speed, low BER and low power consumption performance of the underwater inductive coupling communication system based on power carrier technology designed in this paper, and also prove the feasibility of applying the power carrier technology to the inductive coupling communication system in order to enhance the signal transmission rate and reduce the signal BER.

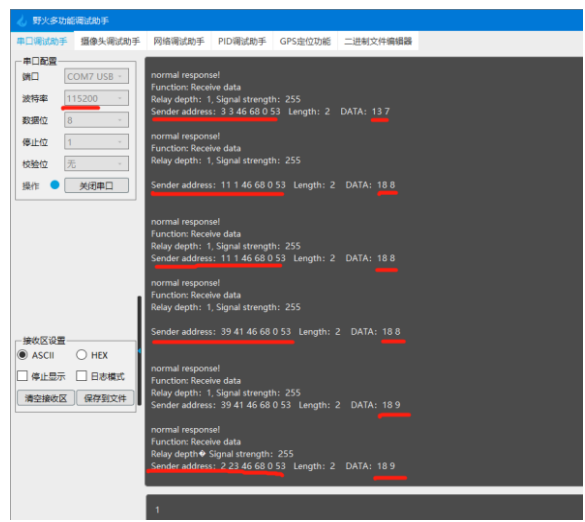

Fig. 19. Function test results of multi-point communication.

## 5 Conclusion

In this paper, power carrier technology was introduced to the underwater inductive coupling communication system to enhance the transmission speed and reduce the bit error rate. The communication channel model was constructed to obtain the optimal parameters of the channel. According to the demand of system communication and power consumption, the system hardware and software design was implemented based on STM32 microprocessor is carried out. Through the modulation and demodulation mode comparison test, DPSK was adopted to the power carrier technology, which demonstrated its superiority in improving the communication quality and reducing the bit error rate. Test results in the laboratory environment show that the system can achieve efficient data communication with a rate of 115200bps in the 700m channel, and the static power consumption is as low as 660μA. The communication and low-power solution proposed in this paper for the underwater inductive coupling communication system realizes the goal of the underwater communication system to have fast and long-term stable operation in the marine environment.

# References

[1] Sdiri A, Pinho J C, Ratanatamskul C. Water resource management for sustainable development[J]. Arabian Journal of Geosciences, 2018, 11: 1-2.

[2] Song X, Du J, Wang S, et al. Research Progress of Marine Scientific Equipment and Development Recommendations in China[J]. Strategic Study of CAE, 2020, 22(6): 76-83.

[3] Dalei S, Gao S, Xu M, et al. Hardware design of a submerged buoy system based on electromagnetic inductive coupling[J]. MATEC Web of Conferences, 2016, 75: 01001.

[4] Du R, Li X, Yang S, et al. A new method for improving the inductively coupled data transmission rate of mooring buoy based on carrier signal frequency selection[J]. Nanotechnology and Precision Engineering, 2020, 3(2): 96-103.

[5] Zu-rongQIU, QiZHANG, Hong-zhiLI, et al. Design of inductive coupling channel analysis system based on LabVIEW[J]. Journal of measurement science and instrumentation, 2018, 9(4): 360-6.

[6] Yu Z, Yingjie L, Yuanhong R, et al. Orthogonal Frequency Division Multiplexing Adaptive Technology for Multinode Users of Seawater Channel Based on Inductively Coupled Mooring Chain[J]. Journal of Ocean University of China, 2023, 22(5): 1243-52.

[7] Yu Z, Chen F, Yan-Fang L U, et al. Frequency Selectivity Analysis Based on Inductively Coupled Channel for Current Transmission Through Seawater[J]. China Ocean Engineering, 2021, 622–630

[8] Kurz A, Neal M. Wireless data transmission between a base station and a transponder via inductive coupling: 20110243258[P]. 2011-06-10.

[9] Soar R J. An inductively coupled power and data transmission system[J]. Energies. 2023; 16(11): 4417

[10] Qi J, Chen X, Liu X. Advances of Research on Low-Voltage Power Line Carrier Communication Technology[J]. Power System Technology, 2010, 34(5): 161-72.

[11] Wu M, Lou J, Wang X. Home network based on LV power line carrier communication technology[J]. Proceedings of the Chinese Society of Universities, 2003, 15(5): 77-82.

[12] Peng D, Pan Y, Tu Z, et al. A Study on the Application of Power Line Carrier Technology to Automotive Data Transmission[J]. Automotive Engineering, 2004, 26(4): 435.

[13] Song D L, Gao S, Xu M, et al. Hardware design of a submerged buoy system based on electromagnetic inductive coupling [C]. Munich, GERMANY: the International Conference on Measurement Instrumentation and Electronics (ICMIE), 2016.

[14] Su Y G, Zhou W, Hu A P, et al. Full-Duplex Communication on the Shared Channel of a Capacitively Coupled Power Transfer System[J]. Ieee Transactions on Power Electronics, 2017, 32(4): 3229-39.